\documentstyle[12pt,aaspp4]{article}
\begin{document}
\input epsf
%
%
\def\etal{{\it et al.}}
\def  \vs          {{\it vs.} }
\def  \eg          {{\rm e.g.}}
\def\G{$\Gamma$}
\def\grs{$\gamma$-rays}
\def\egs{{\it EGRET}~}
\def\gr{$\gamma$-ray}
\def\IUE{{\it IUE}}

%
%
\def  \La          {\ifmmode {\rm Ly}\alpha \else Ly$\alpha$\fi}
\def  \Ka          {\ifmmode {\rm K}\alpha \else K$\alpha$\fi}
\def  \Kb          {\ifmmode {\rm K}\beta \else K$\beta$\fi}
\def  \Lb          {\ifmmode {\rm L}\beta \else L$\beta$\fi}
\def  \Ha          {\ifmmode {\rm H}\alpha \else H$\alpha$\fi}
\def  \Hb          {\ifmmode {\rm H}\beta \else H$\beta$\fi}
\def  \Pa          {\ifmmode {\rm P}\alpha \else P$\alpha$\fi}
\def  \HeI        {\ifmmode {\rm He}\,{\sc i}\,\lambda5876
                     \else He\,{\sc i}\,$\lambda5876$\fi}
\def  \HeII        {\ifmmode {\rm He}\,{\sc ii}\,\lambda4686
                     \else He\,{\sc ii}\,$\lambda4686$\fi}
\def  \HeIIa        {\ifmmode {\rm He}\,{\sc ii}\,\lambda1640
                     \else He\,{\sc ii}\,$\lambda1640$\fi}
\def  \HeIIb        {\ifmmode {\rm He}\,{\sc ii}\,\lambda1085
                     \else He\,{\sc ii}\,$\lambda1085$\fi}
\def  \CII        {\ifmmode {\rm C}\,{\sc ii}\,\lambda1335
                     \else C\,{\sc ii}\,$\lambda1335$\fi}
\def  \CIII        {\ifmmode {\rm C}\,{\sc iii}\,\lambda977
                     \else C\,{\sc iii}\,$\lambda977$\fi}
\def  \CIIIb       {\ifmmode {\rm C}\,{\sc iii]}\,\lambda1909
                     \else C\,{\sc iii]}\,$\lambda1909$\fi}
\def  \CIV         {\ifmmode {\rm C}\,{\sc iv}\,\lambda1549
                     \else C\,{\sc iv}\,$\lambda1549$\fi}
\def  \bOIIIb       {\ifmmode {\rm [O}\,{\sc iii]}\,\lambda5007
                     \else [O\,{\sc iii]}\,$\lambda5007$\fi}
\def  \OIIIb       {\ifmmode {\rm O}\,{\sc iii]}\,\lambda1663
                     \else O\,{\sc iii]}\,$\lambda1663$\fi}
\def  \OVIII        {\ifmmode {\rm O}\,{\sc viii}\,653~{\rm eV}
                     \else O\,{\sc viii}\,$653~{\rm eV}$\fi}
\def  \OVII        {\ifmmode {\rm O}\,{\sc vii}\,568~eV
                     \else O\,{\sc vii}\,$568~{\rm eV}$\fi}
\def  \OVI         {\ifmmode {\rm O}\,{\sc vi}\,\lambda1035
                     \else O\,{\sc vi}\,$\lambda1035$\fi}
\def  \OIVb         {\ifmmode {\rm O}\,{\sc iv]}\,\lambda1402
                     \else O\,{\sc iv]}\,$\lambda1402$\fi}
\def  \bOIIb       {\ifmmode {\rm [O}\,{\sc ii]}\,\lambda3727
                     \else [O\,{\sc ii]}\,$\lambda3727$\fi}
\def  \bOIb       {\ifmmode {\rm [O}\,{\sc i]}\,\lambda6300
                     \else [O\,{\sc i]}\,$\lambda6300$\fi}
\def  \OI       {\ifmmode {\rm [O}\,{\sc i]}\,\lambda1304
                     \else [O\,{\sc i]}\,$\lambda1304$\fi}
\def  \NII         {\ifmmode {\rm N}\,{\sc ii}\,\lambda1084
                     \else N\,{\sc ii}\,$\lambda1084$\fi}
\def  \NIII         {\ifmmode {\rm N}\,{\sc iii}\,\lambda990
                     \else N\,{\sc iii}\,$\lambda990$\fi}
\def  \NIIIb         {\ifmmode {\rm N}\,{\sc iii]}\,\lambda1750
                     \else N\,{\sc iii]}\,$\lambda1750$\fi}
\def  \NIVb         {\ifmmode {\rm N}\,{\sc iv]}\,\lambda1486
                     \else N\,{\sc iv]}\,$\lambda1486$\fi}
\def  \NV          {\ifmmode {\rm N}\,{\sc v}\,\lambda1240
                     \else N\,{\sc v}\,$\lambda1240$\fi}
\def  \MgII        {\ifmmode {\rm Mg}\,{\sc ii}\,\lambda2798
                       \else Mg\,{\sc ii}\,$\lambda2798$\fi}
\def  \CVI        {\ifmmode {\rm C}\,{\sc vi}\,368~eV
                       \else C\,{\sc vi}\,368~eV\fi}
\def  \SiIV         {\ifmmode {\rm Si}\,{\sc iv}\,\lambda1397
                     \else Si\,{\sc iv}\,$\lambda1397$\fi}
\def  \bFeXb       {\ifmmode {\rm [Fe}\,{\sc x]}\,\lambda6734
                       \else [Fe\,{\sc x]}\,$\lambda6734$\fi}
\def  \MgX        {\ifmmode {\rm Mg}\,{\sc x}\,\lambda615
                       \else Mg\,{\sc x}\,$\lambda615$\fi}
\def  \MgXI        {\ifmmode {\rm Mg}\,{\sc xi}\,1.34~keV
                       \else Mg\,{\sc xi}\,1.34~keV\fi}
\def  \MgXII      {\ifmmode {\rm Mg}\,{\sc xii}\,1.47~keV
                     \else Mg\,{\sc xii}\,1.47~keV\fi}
\def  \bNeVb      {\ifmmode {\rm [Ne}\,{\sc v]}\,\lambda3426
                     \else [Ne\,{\sc v]}\,$\lambda3426$\fi}
\def  \NeVIII      {\ifmmode {\rm Ne}\,{\sc viii}\,\lambda774
                     \else Ne\,{\sc viii}\,$\lambda774$\fi}
\def  \SiVIIa      {\ifmmode {\rm Si}\,{\sc vii}\,\lambda70
                     \else Si\,{\sc vii}\,$\lambda70$\fi}
\def  \NeVIIa      {\ifmmode {\rm Ne}\,{\sc vii}\,\lambda88
                     \else Ne\,{\sc vii}\,$\lambda88$\fi}
\def  \NeVIIIa      {\ifmmode {\rm Ne}\,{\sc viii}\,\lambda88
                     \else Ne\,{\sc viii}\,$\lambda88$\fi}
\def  \NeIX      {\ifmmode {\rm Ne}\,{\sc ix}\,915~eV
                     \else Ne\,{\sc ix}\,915~eV\fi}
\def  \NeX      {\ifmmode {\rm Ne}\,{\sc x}\,1.02~keV
                     \else Ne\,{\sc x}\,1.02~keV\fi}
\def  \SiXII        {\ifmmode {\rm Si}\,{\sc xii}\,\lambda506
                       \else Si\,{\sc xii}\,$\lambda506$\fi}
\def  \SiXIII      {\ifmmode {\rm Si}\,{\sc xiii}\,1.85~keV
                     \else Si\,{\sc xiii}\,1.85~keV\fi}
\def  \SiXIV      {\ifmmode {\rm Si}\,{\sc xiv}\,2.0~keV
                     \else Si\,{\sc xiv}\,2.0~keV\fi}
\def  \SXV      {\ifmmode {\rm S}\,{\sc xv}\,2.45~keV
                     \else S\,{\sc xv}\,2.45~keV\fi}
\def  \SXVI      {\ifmmode {\rm S}\,{\sc xvi}\,2.62~keV
                     \else S\,{\sc xvi}\,2.62~keV\fi}
\def  \ArXVII      {\ifmmode {\rm Ar}\,{\sc xvii}\,3.10~keV
                     \else Ar\,{\sc xvii}\,3.10~keV\fi}
\def  \ArXVIII      {\ifmmode {\rm Ar}\,{\sc xviii}\,3.30~keV
                     \else Ar\,{\sc xviii}\,3.30~keV\fi}
\def  \FeI_XVI      {\ifmmode {\rm Fe}\,{\sc 1-16}\,6.4~keV
                     \else Fe\,{\sc 1-16}\,6.4~keV\fi}
\def  \FeXVII_XXIII     {\ifmmode {\rm Fe}\,{\sc 17-23}\,6.5~keV
                     \else Fe\,{\sc 17-23}\,6.5~keV\fi}
\def  \FeXXV      {\ifmmode {\rm Fe}\,{\sc xxv}\,6.7~keV
                     \else Fe\,{\sc xxv}\,6.7~keV\fi}
\def  \FeXXVI      {\ifmmode {\rm Fe}\,{\sc xxvi}\,6.96~keV
                     \else Fe\,{\sc xxvi}\,6.96~keV\fi}
\def  \FeLa     {\ifmmode {\rm Fe}\,{\sc L}\,0.7-0.8~keV
                     \else Fe\,{\sc L}\,0.7-0.8~keV\fi}
\def  \FeLb     {\ifmmode {\rm Fe}\,{\sc L}\,1.03-1.15~keV
                     \else Fe\,{\sc L}\,1.03-1.15~keV\fi}
%
%
\def  \L46         {$ L_{46} $}
\def  \Mo          {$ M_{\odot} $}     
\def  \Lo          {$ L_{\odot} $}     
\def  \Fn          {$ F_{\nu} $}
\def  \Ln          {$ L_{\nu} $}
\def  \LEdd        {$ L_{\rm Edd} $}    
\def  \MBH         {$ M_{\rm BH} $}     
\def  \m9          {$ m_9 $}
\def  \mdot        {$ \dot{m} $}        
\def  \Mdot        {$ \dot{M} $}        
\def  \Te          {$ T_{\rm e} $}      
\def  \Tex         {$ T_{\rm ex} $}     
\def  \Teff        {$ T_{\rm eff} $}    
\def  \THIM        {$ T_{\rm HIM} $}    
\def  \TC          {$ T_{\rm C} $}      
\def  \Ne          {$ N_{\rm e} $}      
\def  \nh          {$ n_{\rm H} $}      
\def  \N10         {$ N_{10} $}
\def  \Ncol        {$ N_{\rm H} $}    
\def  \rin         {$ r_{\rm in} $}
\def  \rout        {$ r_{\rm out} $}
\def  \rav         {$ r_{\rm av} $}
\def  \jc          {$ j_c(r) $}
\def  \Rc          {$ R_c(r) $}
\def  \Ac          {$ A_c(r) $}
\def  \nc          {$ n_c(r) $}
\def  \Cr          {$ C(r) $}
\def  \Rf          {$ \rm R_f $}     
\def  \Cf          {$ \rm C_f $}     
\def  \El          {$ E_l(r) $}
\def  \el          {$ \epsilon_l(r) $}
\def  \PSI         {$ \Psi (t) $}
\def  \me          {$ m_{\rm e} $}            
\def  \mp          {$ m_{\rm p} $}            
\def  \rBLR        {$ r_{\rm BLR} $}          
\def  \rNLR        {$ r_{\rm NLR} $}          
\def  \aox         {$ \alpha_{ox} $}          %
\def  \Ux          {$ \rm U_{\rm x} $}  
%
%
\def  \kms         {\hbox{km s$^{-1}$}}          
\def  \ergs        {\hbox{erg s$^{-1}$}}              
\def  \ergsHz      {\hbox{erg s$^{-1}$ Hz$^{-1}$}}   
\def  \cc          {\hbox{cm$^{-3}$}}
\def  \cmii        {\hbox{cm$^{-2}$}}
\def  \cms         {\hbox{cm s$^{-1}$}}      
\def  \mic         {$\mu$m}
\def  \vs          {{\it vs.} }
\def  \etal        {{\it et al.}}
%
%
\newbox\grsign \setbox\grsign=\hbox{$>$} \newdimen\grdimen \grdimen=\ht\grsign
\newbox\simlessbox \newbox\simgreatbox \newbox\simpropbox
\setbox\simgreatbox=\hbox{\raise.5ex\hbox{$>$}\llap
     {\lower.5ex\hbox{$\sim$}}}\ht1=\grdimen\dp1=0pt
\setbox\simlessbox=\hbox{\raise.5ex\hbox{$<$}\llap
     {\lower.5ex\hbox{$\sim$}}}\ht2=\grdimen\dp2=0pt
\setbox\simpropbox=\hbox{\raise.5ex\hbox{$\propto$}\llap
     {\lower.5ex\hbox{$\sim$}}}\ht2=\grdimen\dp2=0pt
\def\simgreat{\mathrel{\copy\simgreatbox}}
\def\simless{\mathrel{\copy\simlessbox}}
\def\simprop{\mathrel{\copy\simpropbox}}
\def \GINGA  {{\it GINGA}}
\def \Ginga  {{\it Ginga}}
\def \ASCA   {{\it  ASCA}}
\def \asca   {{\it  ASCA}}
\def \BBXRT  {{\it BBXRT}}
\def \ROSAT  {{\it ROSAT}}
\def \XTE    {{\it XTE}}
\def \XLi    {XL$_1$}
\def \XLii   {XL$_2$}
\def \XLiii  {XL$_3$}
\def \XMi    {XM$_1$}
\def \XMii   {XM$_2$}
\def \XMiii  {XM$_3$}
%
%
%
\def \aap     { {\it Astr. Ap.}, }
\def \aaps    { {\it Astr. Ap. Suppl.}, }
\def \aj      { {\it A. J.}, }
\def \apj      { {\it Ap. J.}, }
\def \apjl     { {\it Ap. J. (Letters)}, }
\def \apjs    { {\it Ap. J. Suppl.}, }
\def \iau      { {\it International Astronomical Union} }
\def \mnras   { {\it M.N.R.A.S.}, }
\def \pasp    { {\it Pub.A.S.P.}, }
\def \BAAS  { {\it Bulletin of the American Astronomical Society}}
\def \Icarus  { {\it Icarus}}
\def \JMolSpec  { {\it J. Molec. Spectrosc.}}
\def \JQSRT  { {\it J. Quant. Spectrosc. Rad. Trans.}}
\def \Science  { {\it Science}}
\def \JGR  { {\it J. Geophys. Res.}}
\def \PhDthesis  { {\it Ph. D. thesis}}
\def \JOptSocAm  { {\it J. Opt. Soc. Am.}}
\title{The Soft X-ray Spectrum of Scattering--Dominated AGN }
\author{Hagai Netzer \altaffilmark{1},      
T.J.,Turner \altaffilmark{2,3},
I.M.,George \altaffilmark{2,3}} 
\altaffiltext{1}{School of Physics and Astronomy and the Wise
Observatory, The Raymond and  Beverly Sackler Faculty of Exact Sciences,
Tel-Aviv University, Tel-Aviv 69978, Israel}
\altaffiltext{2}{Laboratory for High Energy Astrophysics,
  Code 660, NASA/Goddard Space Flight Center, Greenbelt, MD 20771}
\altaffiltext{3}{University Space Research Association}

\begin{abstract}
This paper discusses the properties of scattering--dominated active galactic
nuclei (AGN). We define these to be AGN for which the direct line-of-sight to
the continuum source is obscured by Compton-thick material. The aim is to
construct, for the first time, a model consistent with X-ray line luminosities,
line ratios and various luminosity indicators. The \ASCA\ spectra of
six such sources  show 
several X-ray lines that  can be reliably measured,
mostly due to highly ionized magnesium, silicon sulphur and iron. These enable
us to investigate the physical
conditions of the scattering material. The sources 
 show evidence of He-like and H-like iron lines that are
likely to be produced in hot (T$\sim 10^6$ K) photoionized gas. By measuring
the EW of the lines, and by constructing a diagnostic line-ratio diagram, we
demonstrate  that the silicon and magnesium lines are produced by the same gas
emitting the highly ionized iron lines. 
The properties of this gas are rather different from the
properties of  warm absorbers in type I AGN. 
Neutral 6.4 keV iron lines are also detected, originating in a
different component which can be either Compton-thin or Compton-thick.
The best measured iron lines suggest an enhancement of  
${\rm n_{Fe}/n_H}$ by a factor $\sim 
2$ compared to solar, in both the hot and cool Compton-thin components.
We further show 
that in four of the sources, the 
Fe \Ka(6.4~keV)/\Hb\ ($\lambda 4861 \AA$) 
line ratio is consistent with that predicted for typical
 narrow line region clouds, and the reddening corrected \Hb\ is known, 
provided the column density is larger than $\sim 10^{22.5}$ \cmii\ , \aox\ is
smaller than 1.3. 
For some sources, this is  a viable alternative to the commonly
assumed Compton thick medium as the origin of
the 6.4 keV iron line.
 
{\it Subject headings:}
 galaxies:abundances - galaxies:Seyfert - galaxies:active
 - line:formation - X-ray:galaxies

\end{abstract}

\section{Introduction: Scattering--Dominated AGN}
The optical and X-ray properties of type II AGN (i.e. those showing prominent
narrow emission lines and very faint, if any, broad lines) have been discussed
in numerous recent  papers that contain the analysis of their morphology,
spectrum, geometry and relationship to type I (broad emission line) AGN.
Various names, including Seyfert 2, narrow emission line galaxies (NELG), and
narrow line X-ray galaxies (NLXGs) have been used to describe these objects.
Obviously, there is some subjectivity in the classification of type II AGN 
leading to the assignment of a more than one ``type'' for some objects 
which have been studied by several authors. 
The geometry of the innermost region of such sources is a fundamental, yet
still an open issue and the reader is referred to Antonucci (1993), and
Mulchaey \etal\ (1994) for discussion and references regarding these questions.

X-ray observations  offer a unique view of type II AGN,  since X-rays can
penetrate large column densities.  This has been a subject of much research, 
for example, see recent papers by 
Turner \etal\ (1997a,b,1998). These authors found some surprising
similarities between the X-ray spectra of type I and type II AGN. In some type
II sources, the equivalent width (EW) of the 6.4 keV line is similar to that
observed in Seyfert 1s and the line profiles show broad, redshifted wings. In
other type II sources, like NGC~1068, EW(Fe \Ka) is an order of magnitude
larger than in type I AGN, suggesting that the line is seen against a reduced
continuum, presumably due to obscuration. Evidently, type II AGN fall into at
least two X-ray  categories; those where the central source is directly
observed below 10 keV, and those where it is not. 
Hereafter we refer to those type II AGN
whose 0.5--10 keV spectra are dominated by scattered radiation,
``scattering--dominated AGN'', and they are the subject of this paper. 
We expect a good, but not necessarily a
one-to-one correlation between such objects and those type II AGN who show
highly polarized continuum and broad optical/UV emission lines, due to
scattering (e.g. Tran 1995).

This paper investigates the properties of scattering--dominated AGN through
detailed analysis of their 0.5--10 keV spectrum. We address the nature of the
scattering medium  and try  to deduce its level of ionization, column density 
and covering fraction. We  also  investigate the metallicity of the gas and
compare its properties to the ionized gas in Seyfert 1 galaxies, and to
the narrow line region (NLR) gas. The analysis is aimed at a small number of
scattering--dominated  AGN whose \ASCA\ spectra are of a sufficiently high
quality to enable the measurement of at least 3 X-ray emission lines. It 
also suggests several new avenues for future study of such sources in
preparation for the coming  {\it AXAF}\ and {\it XMM}\ missions.
In \S2 we discuss the predicted spectra of such sources. 
In \S3 we compare predictions to a detailed analysis of the \ASCA\ spectra 
of six such galaxies. In 
\S4 we discuss several implications of such a comparison, 
and implications for the state and the location of 
the scattering medium and its composition.

\section{The X-ray spectra of scattering--dominated AGN}

In this section we  show new calculations pertaining to
scattering--dominated AGN. The calculations assume a simple geometry
of a point-like source of X-ray and UV radiation surrounded by gas clouds of
various locations, density and column density. The gas is photoionized by the
radiation of the central source and no other excitation is assumed. 
The geometry is
such that the line-of-sight to the central source is completely covered by a
column density of 10$^{24}$ \cmii\ or larger but the scattering medium
has a clear view of the center and we have a clear view of the scattering gas.
The  parameters of the model are
the gas covering factor, \Cf=$\Omega /4 \pi$, where $\Omega$ is the solid
angle subtended at central source,  density \nh, column density \Ncol, and the gas 
location. 
We shall make extensive use of the ``X-ray
Ionization Parameter'', Ux=${\rm Q_{0.1-10~keV}}/{\rm n_H}c$ (Netzer 1996), 
where ${\rm Q_{0.1-10~keV}}$ is the incident photon flux in the 
0.1--10 keV band. This parameter 
is closely related to the ionization level of most observed highly ionized
species. Conversion factors between Ux and the U$_{13.6~ev}$ ionization parameter
(defined over the entire Lyman continuum), are given in George \etal\ (1998).

The scattering gas is assumed
to be distributed in a shell geometry and is made of a large number of small
clouds with dimensions similar to the shell thickness. The shell is not 
complete since the obscurer (possibly a torus?) subtends a non-negligible solid
angle as viewed from the center. For simplicity, we assume
that  Ux and \Ncol\ are the same for all clouds. 
All scattering and
emission processes are assumed to be isotropic and the photons emitted
from the cloud and reaching the observer are
both those which 'leak' between the clouds and those transmitted through the 
clouds. For a uniform distribution of small clouds, the relative fraction
of the two (i.e. the leaking photons and the attenuated ones) is uniquly
specified by the covering factor \Cf. The real situation can be rather
more complex. For example, a geometry of \Cf$<1$ may involve extra 
absorption of {\it all} emitted photons. This can happen if a single large
cloud is situated just outside the main scattering region, on the line of
sight. In such a case, all emitted and scattered photons suffer extra 
attenuation on the
way out. The present calculations do not include such complications.

The calculations presented here were performed using the photoionization
code \verb+ION97+. This code includes a full
treatment of all important IR, optical, UV and X-ray transitions and solves,
simultaneously, for the level
of ionization, temperature and opacity of the exposed gas. 
For a full description and more references see Netzer (1996). 
The calculations assume two ``canonical'' spectral energy distributions (SEDs).
We describe those either by the energy slope $\alpha$, i.e. where the 
luminosity, L$_{\rm E}$, is related to the energy, E, as 
(L$_{\rm E} \propto {\rm E}^{-\alpha}$) or the photon slope 
where the number of photons, N, at energy E is given by $\Gamma$
(N$_{\rm E} \propto {\rm E}^{-\Gamma}$).
The first  has 
a UV bump centered at 40 eV, a soft (E$>$100 eV) X-ray component  of 
$\alpha=2$ 
 changing gradually to a hard X-ray slope of $\alpha=0.9$. The ratio of the 
two  continuum components at 1~keV is soft(X-ray)/hard(X-ray)=0.5. For 
this SED, \aox=1.38.  The second SED 
has a similar UV bump but a much flatter hard X-ray slope of $\alpha=0.5$. 
In this case \aox=1.13, typical of low luminosity Seyfert 1 galaxies. 
The fiducial gas density is ${\rm n_H} = 10^8$ \cc\ which adequately represents 
 the physical conditions over the density range of 10$^5$ to 10$^{11}$\cc. 
Unless otherwised specified, the following standard metallicity (which
is similar to solar metallicity) is assumed throughout:\\
H:He:C:N:O:Ne:Mg:Si:S:Ar:Fe=10$^4$:1000:3.7:1.1:8:1.1:0.37:0.35:0.16:0.037:0.4.

The more important ingredients of the model 
are the ionization parameter Ux, which determines the level of ionization
and line intensity, the column density \Ncol, which determines
the fraction of the incident radiation absorbed by individual clouds, and a
parameter which we name ``reprocessing factor'', \Rf(E). This 
energy-dependent parameter, represents 
the fraction of the radiation of the central source 
reaching the observer via scattering. For small \Ncol\ and small \Cf,
it is almost energy independent and is given by \Rf$\simeq \tau_e$\Cf, where
$\tau_e$ is the Compton optical depth of the clouds.

Our aim in this work is to investigate the parameter space 
occupied by scattering-dominated AGN. Thus, the observed continuum must be
consistent with the assumed intrinsic SED, the observed line ratios must
correctly reflect the various neutral and ionized  components, and
\Rf\ must be consistent with the bolometric luminosity of the
source.
Previous studies have only considered a sub-set of these variables.
Here we consider them all self-consistently
for the first time.
Below we discuss the predicted X-ray lines and scattered continuum.

\subsection{X-ray emission lines}
 
\subsubsection{The iron K lines}

The absorption of the hard X-rays by iron ions results in Fe \Ka\
line emission due to
fluorescence (all ions up to Fe{\sc xxiv}), resonance line absorption (ions
with ten or fewer electrons) and recombination. 
Fe \Kb\ transitions, with about 10 percent the
intensity of the corresponding Fe \Ka\ lines, are also emitted.
These processes have been discussed, extensively,
in the literature and will not be reviewed here. The reader is referred to
Kallman and Krolik (1987),  George and Fabian (1991) and Leahy and Creighton 
(1993) for calculations of the Fe \Ka\ lines in
neutral gases, and to Matt (1994) and Matt, Brandt \& Fabian (1996),
for cases involving ionized material. \verb+ION97+ includes all such processes
and the lines are calculated under the assumption of complete isotropy of the 
scattered radiation. This simplified assumption can
introduce  small differences compared with the full, angle dependent 
calculations 
(e.g. Matt \etal\ 1996, Fig. 5). However, the geometry under consideration is sufficiently complex, and most likely involves a combination of various angles,
that we see no reason to consider this additional complication.

The observed EWs depend on the level of ionization, the
column density, the gas expansion or microturbulent motion (which determine
the resonance line optical depth), the incident continuum shape and the iron
 abundance.
Unless otherwise specified, all  calculations  assume
microturbulent motion with a line width  fixed at the local sound speed,
i.e. similar to the hydrogen thermal width.
The calculations are restricted to $\tau_e < 1$  since the present transfer
approximation  is not adequate  for  Compton-thick material.

Fig. 1 shows calculated EW(\Ka) for ``neutral'' (Fe{\sc i}--{\sc xvi}) 
and He-like (Fe{\sc xxv}) iron lines, over a range of ionization parameter and
 column density.  
It demonstrates that for large \Ncol, the maximum EW of the two is comparable.
The reason is that both the scattered continuum and the E$>7.1$ keV flux
are proportional to the column density of the ionized gas, and the absorption
cross sections of the different iron ions are very similar. (The neutral lines
can absorb slightly more photons but the recombination efficiency of
the Fe{\sc xxv} line is larger than the fluorescence yield of the low ionization
ions which compensates for this). This is not the case 
for small \Ncol, where EW(Fe{\sc xxv}) is considerably larger due to 
resonance absorption of continuum photons by the 6.7 keV lines.
As to the dependence on \Cf, see below.

Fig. 2 shows a detailed dependence of the various \Ka\ intensities on Ux,
for the specific column density of 10$^{22.4}$ \cmii and \Cf = 0.5.
The equivalent widths shown are measured with respect to the {\it scattered
continuum} and are designated EW(gas).

As seen from the diagram, {\it the summed}  EW(\Ka) is almost independent of Ux, 
demonstrating that for the assumed gas motion, with the sound-speed
line width, resonance absorption is not 
very important compared with continuum (otherwise the EWs of the iron lines where
 this process can take place,
 would be much larger).
 The reason is the large optical depth in all
resonance transitions. For example, the optical depth of 
the \FeXXV\ line for this \Ncol\ and for Ux=10 is about 9, i.e. 
resonance absorption is efficient only for  
  turbulent motion exceeding about 1000 km/sec.

\subsubsection{The soft X-ray lines}

We define these transitions to include all lines with $0.5<$E$<$6.4 eV. The
strongest are  He-like and H-like lines of the most abundant elements. Under
the conditions assumed here, most of the emitted flux is due to recombination
and resonance absorption is of little significance. 
Auger excited  \Ka\
transitions in ions with more than three electrons can be important too,
despite the very small fluorescence yield, because of the larger photon flux
at low energies (Netzer and Turner 1997). 
Collisionally dominated lines are
usually very weak, because of the low temperature of the photoionized gas.
This is also true for the  collisionally-excited Fe L-shell lines. However,
the recombination lines due to 
Fe L-shell transitions do make significant  contributions to the
observed  flux around 0.8--1.2 keV. 
For the treatment of those lines see Netzer (1996).

Fig. 2 shows also the calculated EW(gas),  as a
 function of Ux, for several soft X-ray lines.
Fig. 3 shows the relative line intensities obtained by normalizing  the  line 
flux to the {\it incident} continuum 
at a specific  energy (0.65 keV). This diagram illustrates the processing efficiency
of the various lines.  Fig. 4 gives yet another view of the information presented
in Fig. 2,  this time over a large
range in column density and ionization parameter.

While most EWs  are not sensitive to the exact value of the covering fraction,
this is not the case when \Cf\ approaches unity. This can be explained by
noting that for optically thin gas, the relative line to scattered continuum
flux is independent of the covering fraction. This is also the case for large
optical depth and \Cf$< 0.9$, since  the line and adjacent continuum photons
undergo similar amount of absorption and much of the observed flux is due to
leakage through ``holes'' in the cloud configuration. However, for large
optical depth and covering fraction approaching unity, much of the line and
continuum radiation gets absorbed on the way out.  Minor differences in the
optical depth structure can result in large changes in equivalent width. The
same is true for situations where the 4$\pi$ covering fraction is much less
than unity but the emitted photons pass, on the way out, through gas with
substantial optical depth. Such situations must be treated with care and are
not considered here. Needless to say, the absolute line flux is very sensitive
to the value of \Cf\ and obtains its maximum value at \Cf$\sim 0.6-0.8$
(Netzer 1996).

\subsection{The scattered continuum}

The SED of the scattered continuum is another indicator of the ionization 
state and  column density of the gas. This 
has been discussed in numerous papers since the first work
of Lightman and White (1988) and many examples 
are shown in Netzer (1993), Netzer, Turner and George (1994), Zycki and 
Czerny (1994), Zycki \etal\ (1994) and Ghisellini, Haardt and Matt (1994).

The 0.5--10 keV continuum of several scattering--dominated AGN (see \S3) are
considerably flatter than typical, type-I AGN continua. We assign this to the
unique chracteristic of scattered SEDs. To illustrate this point, and to
enable a meaningful comparison with the emission line spectrum  (\S4),  we
show, in Fig. 5, two calculations of scattered  spectra, using our canonical
L$_{\rm E} \propto {\rm E}^{-0.9}$ continuum. One has a column density of
10$^{23}$\cmii\ and very low ionization (Ux=10$^{-3}$) and the other a column
of 10$^{22}$\cmii\ and a much higher ionization (Ux=10). The two have the
same covering factor and are combined
in such a way that the relative intensity of the neutral and highly ionized
\Ka\ lines are similar to what is observed in three of the sources under
discussion (\S3).
As evident from the diagram, large-column, low-ionization gas results in a 
rising SED while the slope of the highly ionized gas resembles the
incident continuum. The combination of the two in the right proportions can 
produce very flat continua, as observed in several scattering--dominated AGN.

Calculations such as the ones shown here can be used to compare 
 the observed continuum shape with the relative emission line strength.
 In particular,
the analysis of the Fe \Ka\ blend is most useful since, for a typical
type-I AGN SED, the transition to He-like and H-like iron occurs at
Ux$\simeq few$ (Fig. 2). The relative strengths of such lines must therefore
tell us the relative proportion of the various scattered continuum components.
Finally, the observed continuum luminosity  is related
to the intrinsic SED through the reprocessing factor \Rf.
As shown by Turner \etal\ (1997b), and explained in detail below, this parameter can be inferred from a comparison
of the X-ray luminosity 
 with luminosity indicators at other wavelengths.

\section{Comparison with {\it ASCA} observations}

We have chosen to analyze six narrow line Seyfert galaxies with the most
prominent EW(Fe \Ka) available in the \ASCA\ archives (as of June 1997):
Mrk~3, NGC~6240, NGC~1068, Circinus A,
Mrk~348 and NGC~4388.
These sources are thus likely to be scattering--dominated AGN. 
Three of the sources (Mrk~3, NGC~6240, NGC~1068) are from the 
Turner \etal\ (1997a) sample, and the results from Circinus A have been 
previously presented by Matt \etal\ (1996).
 NGC~1068 has also recently been discussed by Netzer and Turner (1997) and we 
use the line measurement given in that paper. 
 For the others, we have remeasured all prominent X-ray lines, by criteria 
explained  below, and some of our results are therefore  
different from those listed in the original references. 
 Other suspected objects of this type (Turner \etal\ 1997b) have spectra 
with much lower S/N and are not suitable for this kind of 
detailed analysis.

The sources presented here were systematically analyzed using the same method 
as described in Turner \etal\ (1997a). 
 We do not use data below 0.6 keV in the spectral fitting because of 
some uncertainty in the \ASCA\ XRT/SIS calibration, rendering 
those data points artificially low. 
 Data from both pairs of SIS and GIS instruments were analyzed simultaneously, 
but with the normalization of each  dataset allowed to vary relative to the 
others (to allow for small uncertainties in the relative flux calibrations 
of the detectors and the different extraction cells employed).
 All fluxes, luminosities and normalizations are tabulated for SIS0 
(with no correction for the point-spread function, or counts lost off the 
edge of the chip, both are small effects in these cases). 
 We utilized the gain-corrected pulse-invariant (PI) channels for
all instrument datasets, and the spectra were binned for the spectral 
analysis such that each channel contained at least 20 counts allowing 
$\chi^2$ minimization techniques.
 Response matrices released on 1994 Nov 9 and 1995 March 6 were used for the 
SIS and GIS data respectively, and the spectral analysis performed using 
\verb+ XSPEC+ (v10.0).

We employ simple models to try to achieve an acceptable 
parameterization of the X-ray continuum shape for each dataset.
The purpose is to allow accurate line measurement and we attach no physical significance to the assume continuum components.
 To do this we first fit the 0.6--10 keV data excluding the 5--7 keV band 
which is dominated by iron K-shell  line photons. 
 We considered a number of continuum models in turn, finding that a double 
power-law model provided a satisfactory parameterization of each source.
 The absorption of the soft X-ray power-law column was constrained to be 
greater than or equal to the Galactic line-of-sight column (hereafter denoted 
$N_{Hgal}$), while the absorption of the hard power-law was independent of 
this, and unconstrained. 
 Fig.~6 shows the SIS data and fitted models to the six 
datasets.
Many X-ray emission lines are evident in these datasets. The 5--7~keV data 
were returned to the fit, and gaussian model components were added to the 
model until we had tested for the presence of all expected lines.

Line widths were constrained to be narrow ($< 3$ eV), and we took the approach 
of adding numerous narrow components to the model, even when too closely 
spaced to be resolved by \ASCA\ with the intent that we 
model all the line flux in the data. 
 However, we tabulate only the resolvable quantities, i.e. we note the sum of 
Fe{\sc x}--{\sc xxiv}, Fe{\sc xxv--xxvi}+ Fe \Kb, Si{\sc xii-xiii} and
S{\sc xiv--xv}.
 The \SiXIV\ line is currently problematic for \ASCA,
as it is at an  energy close to where most of the calibration `adjustments' 
have been made. 
 Bearing this in mind, the strength of this line is currently subject to 
additional unquantifiable uncertainty, in excess of the statistical error 
tabulated. 

 As this analysis also revealed evidence for some  broad unmodeled residuals 
in the Fe \Ka\ regime, we allowed a broad Fe \Ka\ component, which may 
physically represent either the Compton scattered wing or, more likely 
(Turner \etal\ 1998), 
the asymmetric component from the putative accretion disk. Furthermore, 
we assumed that scattering--dominated AGN have an intrinsic broad Fe \Ka\ line 
component  of the same strength 
and shape as that observed in Seyfert 1 galaxies (Nandra \etal\ 1997, 
Turner \etal\ 1998). This line is scattered, like the
6.4 keV continuum, by the surrounding gas and ought to be included in the fit. 
 This component was constrained to be $\leq 300$~eV, relative to the 
hard X-ray continuum, to keep it within the range of strengths 
observed for Seyfert 1 galaxies. Its inclusion generally improved the fit 
significantly, thus the list in Table~1 takes the 
broad component into account, even 
though it is not tabulated.

Our continuum fitting procedure includes two power-law components, specified
by the photon slope $\Gamma$,  and one or two
 absorbers. The first absorber  
is  the deduced galactic column. The second is required for 
NGC~6240 and Mkn~348, where we detect the direct component at high
energies.  As already mentioned, the only purpose is
 to get a reasonable fit to allow accurate line measurements and 
reliable  estimates of the 2--10 keV luminosities (attenuation was taken into
account when estimating the intrinsic luminosities).
 The  parameters are listed in Table 2 and the ones required for 
  NGC~1068 are given in Netzer and
 Turner (1997) (for this source $L_{2-10}=1.8 \times 10^{41} {\rm erg\ s}^{-1}$).

\section{Discussion}

The following analysis is based on the line fluxes listed in Table 1
as well as on the analysis of the \ASCA\ spectrum of NGC~1068 (Netzer and Turner, 1997,
Table 1). Obviously, the number of objects, and the number of measurable lines per
object, are very small and the information content regarding the group properties
is rather limited. 

The most severe complication in analyzing the \ASCA\ spectra is  the likely
contamination of the nuclear spectrum by extended, non-nuclear emission.
 This may be the result of hot gas in star forming regions, supernova remnants, 
or any other gas at T$\simeq 10^7$~K. 
 The limited \ASCA\ spatial resolution (with a half-power diameter 
$\sim3$~arcmin), combined with the relative weakness of the scattered X-ray 
continuum, makes it almost impossible to separate the photoionized gas and hot
plasma contributions.  
 Indeed, some sources (e.g. NGC~1068) show clear indication of extended
X-ray emission which is most likely due to starburst activity. This issue will
not be resolved before {\it AXAF} observations  
(with a half-power diameter $<$1~arcsec).
 Below we comment on the information obtained from the study of the Fe \Ka\ 
complex and address the potential use of diagnostic diagrams in the
analysis of the spectrum of scattering--dominated AGN.

\subsection{The \Ka\ complex and the iron abundance}

Except for NGC~1068, all our measurements of the 6--7 keV complex are 
somewhat ambiguous since we can not reliably resolve the highly ionized 
(6.7 and 6.96 keV) Fe \Ka\ lines from the neutral \Kb\ ($\sim 7.1$ keV) line.
 Therefore, the analysis of the high ionization lines pertains to the 
{\it combined intensity} of the 
H-like and He-like iron lines, which was obtained by subtracting 
the expected \Kb\ flux  (10\% the flux of the 6.4 keV \Ka\ line) from the total.
This makes the combined Fe{\sc xxv-xxvi} line intensity in NGC~3488 consistent
with zero because of the uncertainty on the \Kb\ flux (see Table 1) and
the ones in Mkn~348 consistent with zero because of the large intrinsic error.
The uncertainty in the relative strength of the 6.4 and 6.9 keV component is
also affected by the uncertainty in the assumed, scattered broad 6.4 keV line.

The galaxies with measurable soft X-ray lines represent 
two different groups. 
 In one object (NGC~1068), the highly ionized iron lines are comparable in 
  strength to the 6.4 keV line.
In three others (Circinus, Mkn~3 and NGC~4388) the combined intensity of the
 He-like and 
H-like iron lines is only about 10-15\% the intensity of the low ionization 
component. Mkn~348 is possibly an intermediate case but the observational
uncertainties are too large to tell.
 The intensity of the high ionization iron lines  in NGC~6240 is unknown, 
because of their  proximity  to the strong 7.1 keV absorption feature due to 
$2 \times 10^{24}$ \cmii\ of neutral absorber. 
 However, a comparison of the 6.4 keV intensity with the soft X-ray lines 
suggests that this source belongs to the same group as Mkn~3 and  Circinus. 
 As for  the  lines of silicon, magnesium and sulphur, there are only three
sources where reliable line ratios can be obtained and they all look quite
similar (see below).

 Regarding the 6.4 keV iron line, a comparison of the  measured line intensities
  with  the calculations shown in 
Figs. 2 and 3, indicate that in all sources where this line is much stronger
than the 6.7--7.0 keV complex, the line  
can not originate in the same component producing the strong magnesium, silicon and sulphur lines.   
 We therefore suggest that in those  sources, much of the 6.4 keV line 
emissivity is due to reprocessing in a large column of low-ionization gas. 
As for  NGC~1068, the 6.4 keV line in this source  may be due to 
warm (T$_e \simeq 2 \times 10^5$~K) photoionized gas (Marshall \etal\ 1993; 
Netzer \& Turner 1997). This idea is in conflict with  the Iwasawa \etal\ (1997) model (see below).
 We conclude that in four (perhaps five) of the six sources, the reprocessing efficiency
(\Rf) of the neutral component is much greater (a factor of 5--10)
 than that of the ionized component.

Given the level of ionization of the gas producing the 6.4 keV iron line, and 
the shape of the ionizing continuum, we can estimate the iron abundance 
from the observed EW(Fe \Ka) in two  
  interesting limits. The first corresponds to Compton--thin gas and 
has been discussed by Krolik and Kallman (1987), Matt \etal\ (1996)
and others. In this case, assuming negligible resonance line absorption and 
complete isotropy of the scattered 
continuum radiation (as appropriate for a case where the scattering medium 
is viewed at all possible angles),
\begin{equation}
{\rm EW(6.4~keV) \simeq 3.17 [\frac {1.11^{1-\Gamma}}{2+\Gamma}] 
    [\frac{F_Y}{0.3}] 
   [\frac{ n_{Fe}/n_H }{ 4 \times 10^{-5} }] \,\, keV } \,\, ,
\end{equation} 
where F$_{\rm Y}$ is the fluorescence yield (of the order of 0.3 for
low ionization iron). In deriving this expression we have adopted the small column density
limit which allows us to neglect the absorption of the emitted 
\Ka\ photons on the
way out. A much slower (logarithmic) dependence on ${\rm n_{Fe}/n_H}$
 is expected at very large
columns.

The second case corresponds to Compton--thick gas, such as the walls of
the hypothetical nuclear torus. This case has recently been
  calculated by Matt \etal\ (1996, 1997) for a range of metallicities and
viewing angles. For a $\Gamma=1.9$ continuum, larger than solar ${\rm n_{Fe}/n_H}$
 and $\cos (i)=0.5$, where $i$ is the
angle between the line of sight and the axis of the torus, Matt \etal\  estimate
(see their Fig. 2),
\begin{equation}
{\rm EW(6.4~keV) \simeq 1.8 
 (1+0.6 \log [\frac{ n_{Fe}/n_H }{ 4 \times 10^{-5} }] ) \,\, keV .}
\end{equation} 
The range of observed angles  can change this value by up to a 
a factor of 1.5.
Thus, for a typical
$\Gamma=1.9$ continuum, 
the Compton--thin and Compton--thick assumptions result in a factor 2 difference
in the estimated iron abundance. The two cases  also predict very different   
6-10 keV continuum shape since the Compton-thick gas produces a much stronger
 7.1 keV absorption edge. Finally,  in  Compton-thick gas, some 10\% of 
the 6.4 keV line 
intensity is in a broad red wing.
 Below we discuss the iron abundance under the two different scenarios.
 While the emphasis is on line intensities,  
we note  that our  continuum
fits do not require the presence of a 7.1 keV absorption in any source 
except NGC~6240. 

The following examples consider two gas components, one represents 
``cold'' (small Ux) gas and the other ``hot'' (large Ux) gas. Each 
scatters the central  radiation and produces a scattered
 continuum. We refer to these as the ``two continuum components''.
 We first assume both components to be Compton-thin.
In estimating the iron abundance from the observed EW, we note that the two continuum components are contributing at 6.4 keV and 
$\Gamma$ is not directly measurable since scattering affects the 
observed continuum shape (\S2.2). 
 We use  the numerical calculations (Fig. 2) and assumed that 
the iron composition is identical in all components.
 We  also note  that resonance absorption is negligible, because iron 
is in a very low ionization state in the cold component and the column density is 
large in the hot component.
 Given these assumptions, we expect each component (hot or cold) to have 
a similar EW(Fe \Ka)  {\it relative to its own continuum}.
 Thus, in those sources  where the 6.7--6.96 keV iron lines are
 much weaker than the
neutral 6.4 keV lines, most of the 6.4 keV continuum is due
to reflection by the neutral component. 
 Given the assumed continuum shape, this implies an iron over abundance
by a factor 2--3 
for Circinus and a factor of 1--1.5 
for Mkn~3.   For NGC~6240 and NGC~4388,   
we estimate the EW relative to the scattered component by 
using our best fit model of these source(Fig. 6). 
 According to the model, about half the observed 6.4 keV continuum is due to 
transmission and the other half due to scattering. We can thus estimate 
EW(6.4 keV) relative to the scattered component and deduce
n$_{\rm Fe}$/\nh$\simeq 1.5-2\times$solar.
 Mkn~348 is so different in this respect that we cannot
reliably estimate EW(Fe \Ka) relative to the scattered continuum.
 The iron abundance in NGC~1068, assuming a Compton-thin gas,
 has been discussed by Marshall \etal\ (1993)
and Netzer \& Turner (1997), and found to be about three times solar. 
Thus, under the Compton-thin assumption,
we have indications of  iron overabundance in 5 sources. 

Regarding the Compton-thick case, the iron abundance inferred from the
observed EW(\Ka\ 6.4 keV) line is about half the value
deduced above,  i.e. consistent with solar for all sources.
As we show below, the analysis of the highly ionized gas enables an independent
 check on the iron composition
because the medium producing such lines is unlikely to be Compton-thick.

\subsection{Diagnostic diagrams}

 Diagnostic diagrams, involving various line ratios, have been used to 
separate Seyfert galaxies from galactic HII regions, and to search for the 
spectroscopic signature of LINERs (e.g. Baldwin, Phillips \& Terlavich, 1981). 
 Below we attempt to use the same method in the X-ray domain, in search for 
the unique signature of scattering--dominated AGN.

There are two major differences between our study and the investigation of
the optical spectrum of LINERs and HII regions. 
 First, the number of available X-ray lines  is very small and we can only 
measure, reliably, three line ratios and construct two such diagrams. 
 Second, given the gas is photoionized, the X-ray line spectrum is dominated by 
recombination and not a single, purely collisionally excited line is 
strong enough to be used. 
 Collisionally excited X-ray lines dominate the spectrum of hot plasmas with
ratios vastly different from that expected in photoionized gas. Thus, there
is no overlap in properties and line ratio diagrams can either be used for
photoionized gas or for hot plasmas.
 In contrast, lower excitation photoionized nebulae contain a mixture
of recombination and collisionally excited lines that provide
 very useful  diagnostics. Thus ratios like \bOIIIb/\Hb\ have been used
to derive the level of ionization of the gas and 
 line ratios involving highly excited O$^{+2}$ transitions have 
been used to investigate the role of shock excited gas in the spectrum of 
LINERs and Seyfert galaxies (e.g. Ferland and Netzer 1983). This kind of analysis
is not yet possible in the X-ray regime. 

Fig. 7 shows a diagnostic diagram composed of the best observed line ratios in
our sample, I(\SiXIII)/I(\FeXXV\ + \FeXXVI) versus 
I(\MgXI)/I(\SiXIII). 
 The first ratio is a good indicator of regions of large ionization parameters, 
with electron temperature of the order of 10$^6$ K, and the second measures the 
conditions in lower ionization gas. 
 Obviously, the excitation and ionization of \MgXI\ and \SiXIII\ is 
rather similar and future analysis, based on oxygen and neon lines, will be 
of greater use. 
 Measurements of the two  ratios are available for  Circinus and NGC~1068 and  
an upper limit can be  obtained for Mrk~3 (see \S2). 
 The diagram shows the location of the three objects along with four 
theoretical curves, this time  assuming 
the L$_{\rm E}\propto {\rm E}^{-0.5}$ continuum. Similar results, with
appropriate scaling of Ux, are obtained for the
 L$_{\rm E}\propto {\rm E}^{-0.9}$ continuum used in most other calculation. 
 The curves are series of increasing Ux for various column
density and gas composition. 
 The  solid lines are standard composition models for three column densities, 
\Ncol=10$^{22.4}$~\cmii, 10$^{23}$~\cmii\ and 10$^{23.3}$~\cmii. 
 The dotted line is for \Ncol=10$^{23}$~\cmii\ but 
with ${\rm n_{Fe}}$/\nh\ three times larger.

Inspection of the line ratio diagram, and the observed spectra, suggest that: 
\begin{enumerate}
\item	The observed line ratios cannot be simultaneously obtained in a 
	single temperature collisionally ionized gas.
	Under such conditions, the temperature required to ionize 
	Fe{\sc xxv} and Fe{\sc xxvi} is inconsistent with the observed 
	strength of the silicon and magnesium lines. 
	If all lines are produced in a single component, this gas must
	be photoionized. 
	Indeed, \ASCA\ spectra of starburst galaxies 
	(e.g. Ptak \etal, 1997) generally show strong silicon and sulphur 
	lines but little or no emission from neutral Fe \Ka. 
\item	The inferred ionization parameter, $U_X\sim 1$, is 3--10 times larger 
	than the ionization parameter of the highly ionized (warm absorber) 
	gas in Seyfert 1 galaxies (George \etal, 1998).  
	This is in agreement with the finding of Turner \etal\ (1997a). 
	We have examined the properties of this  gas and found a mean electron
	temperature of  about 10$^6$ K and no noticeable soft absorption 
	features.  
	Such gas, on the line of sight to a typical AGN continuum, with a 
	column density not exceeding 10$^{23}$ \cmii\ (the column density of 
	the great majority of warm absorbers in Seyfert 1 galaxies,
	see George \etal\ 1998) would escape detection by \ASCA\ type 
	instruments. 
\item	The diagnostic diagram cannot, by itself, be used to infer the 
	Fe/Si abundance ratio since large column densities mimic the 
	appearance of a small-column with large Fe/Si. 
	As argued in \S4.1, the analysis of the 6.4 keV lines suggest large 
	${\rm n_{Fe}}$/\nh\ in all sources if the emitting medium is Compton-thin. 
	If  the high ionization components have similar compositions, then 
	according to the diagram, they must have relatively small column 
	densities, perhaps similar to the lowest column shown in Fig. 7.
\item 	The weakness of \MgXII\ and \SiXIV\ lines is somewhat 
	surprising. 
	The lines are predicted to be similar in strength to the lower 
	ionization magnesium and silicon lines (Figs. 2 and 3) yet we could 
	only obtain upper limits. 
	The difficulty of detecting the \SiXIV\ line  may partly be explained by 
	the notorious detector/mirror features in \ASCA\ around 2 keV.
\end{enumerate}

\subsection{The scatterer location and the value of \Rf}

So far we have focused on relative line intensities and line-to-continuum 
flux ratios. 
 These  are useful in determining the ionization and composition but do not 
reveal the reprocessing efficiency, \Rf, since EW(Fe 6.4 keV) is almost 
independent of the column density (Fig. 1).  
 \Rf\ can not be obtained by comparing the absolute  
line flux with the intrinsic  luminosity since the latter 
is not known.  
 However, there are several other luminosity indicators at longer 
wavelength, including the infrared flux and the \bOIIIb\ luminosity 
(e.g. Mulchaey \etal, 1994, see extensive discussion in Turner \etal, 1997b,1998),
that can be used. 
 Here we chose to use the reddening-corrected narrow \Hb\ line as our 
luminosity indicator. 
 This  is  similar to the L(\bOIIIb) method used in Turner \etal\ but 
enables a more direct comparison with the intrinsic ultraviolet luminosity.

Measurements of the narrow \Hb\ flux for the six sources,
as well as for most
known type II AGN, are readily available (Mulchaey \etal, 1994; Polletta
\etal, 1996; and references therein). 
 The above references contain also the measured \Ha/\Hb\ line ratio which we 
use to correct for reddening and to obtain the intrinsic \Hb\ flux. 
 In the following we assume  an intrinsic I(\Ha)/I(\Hb)=3.0 and a simple 
galactic type reddening with  A$_{\rm V}$/E$_{\rm B-V}$=3.1. 
Reddening corrected \Hb\ fluxes obtained this way are listed in Table 1.

A word of caution is in order.
Applying a simple, screen-type 
 reddening correction to the spectrum of Circinus and 
NGC~6240 is problematic since the observed Balmer decrement in both galaxies
is very large 
(e.g. Fosbury and Wall 1979 for the case of NGC~6240). 
 In addition, much of the \Hb\ flux in NGC~6240 is likely due to luminous 
star--forming regions in this galaxy.
 In both cases, and probably in many other narrow--line galaxies showing
large Balmer decrements, the geometry 
is  rather complex with several clouds along each line--of--sight. We may be
looking into  dusty environments for which a simple 
correction factor is inappropriate.   
 This can  invalidate the  reddening-corrected \Hb\ fluxes used here.

The theoretical I(Fe \Ka)/I(\Hb) is easily obtained from the spectral energy 
distribution, the column density and the iron abundance. 
 For low ionization, small Balmer optical depth gas, 
the number of \Hb\ photons is a known fraction 
(about 0.12  for Case B recombination) of the Lyman continuum photon flux and 
the number of Fe \Ka\ photons is a known fraction of the ionizing E$>7.1$ keV 
flux. The case of interest for this study is gas with very large Lyman
optical depth yet relatively small hard X-ray  optical depth. 
 Defining Q$_{7.1~keV}$ as the photon flux above 7.1 keV, and Q$_{13.6~eV}$ 
as the Lyman continuum photon flux, we get for this case
\begin{equation}
{\rm I(Fe \Ka)/I(\Hb)} \simeq 1.5 \times 10^3 \exp(- \tau_{\rm 6.4 keV}) 
  [\frac{ {\rm Q_{7.1~keV}}}{ {\rm Q_{13.6~eV}}} ] 
   [1-\exp(-\tau_{\rm 7.1 keV})] \,\, ,
\end{equation}
where 
\begin{equation}
\tau_{\rm 7.1 keV} \simeq 0.13 [\frac{ {\rm N_{col}} }{ 10^{23} } ]
      \frac { ({\rm n_{Fe}/n_H)} }{ 4 \times 10^{-5} } \,\, ,
\end{equation}
is the 7.1 keV optical depth due to iron (assumed to be much smaller than
unity), and
$\tau_{\rm 6.4 keV}$ is the  absorption optical depth due to all metals,
 at 6.4 keV (of the same 
order as $\tau_{\rm 7.1 keV}$ for solar ${\rm n_{Fe}/n_H}$). Thus at small $\tau_{\rm 7.1 keV}$,
the line ratio increases with the iron abundance.
Major complications arise due to  absorption of the
E$>7.1$ keV photons by elements other than iron and by the non-negligible opacity
at 6.4 keV, resulting in the destruction of emitted \Ka\ photons. 
 This  makes the Fe \Ka\ emissivity sensitive to the covering fraction since 
the \Ka\ photons emitted by one cloud can be absorbed by another cloud. 
 Having in mind the NLR conditions, we assume in the following  \Cf=0.1.

Fig.~8 shows a series of calculated I(Fe \Ka)/I(\Hb) for
 N$_{\rm E} \propto {\rm E}^{-1.9}$
 continuum, solar metallicity, \nh=10$^4$\cc\ applicable to the NLR,
 and various  \aox. 
 To enable a comparison with the expressions given above, we note that for 
those models with \aox=1.3,  ${\rm Q_{0.1-10~keV}}/{\rm Q_{7.1~keV}} = 130$ and  
${\rm Q_{13.6~eV}}/{\rm Q_{0.1-10~keV}}=52$.

The diagram shows that for \Ux\ $=10^{-3}$ and 
\Ncol$>10^{21.7}$ \cmii,   \Hb\  is already emitted at maximum efficiency while
the Fe \Ka\ flux is proportional to the column density. 
 For \Ncol$>10^{23.5}$ \cmii, both lines are emitted at maximum efficiency 
and their ratio reflects the spectral energy distribution, ${\rm n_{Fe}/n_H}$
and the destruction of the Fe \Ka\ photons. 

Inspection of Fig.~8  and  Table 1 suggests that in four sources, 
NGC~1068,  NGC~6240  Mrk~3 and Mkn~348, the I(Ka)/I(\Hb) line
 ratio is below 0.2, which is 
consistent with  solar ${\rm n_{Fe}/n_H}$  for 
\Ncol$< 10^{23.5}$ \cmii\ for \aox=1.1. 
 Assuming an iron overabundance by 2--3 reduces the required column to
below 10$^{23}$ \cmii\ for the same \aox. Furthermore, in three of the four
cases the required column can be substantially smaller than the above
mentioned upper limit.  
 The  column density is smaller if the UV bump is weaker than assumed and
larger if \aox\ is larger than assumed.  
 Thus in about half the sources the 6.4 keV iron line 
 could originate in the NLR 
if the clouds in that region have column densities 
exceeding about 10$^{22.5}$ \cmii. 
  Circinus and NGC~4388 are different since  I(Fe \Ka)/I(\Hb) in those sources 
is an order of magnitude larger than in the other three.
 As shown in the diagram, this is unlikely to be due to a much larger column density. 
 Either the X-ray source is very bright compared with the UV source (very small
\aox) or else there is an additional, large covering fraction, 
\Ka\ producing component
 which is neutral, very thick and inefficient \Hb\ emitter.

Current NLR models (see Ferguson \etal\ 1997 and references therein) do
not make definite predictions regarding the size of the NLR clouds, since most 
observed narrow
lines originate in  the highly ionized, H{\sc ii} part of the clouds. 
 An obvious complication is if the NLR gas is dusty (see Netzer and Laor 1993 
and references therein).  
 For example, it is conceivable that the NLR clouds are the illuminated faces 
of  dusty molecular clouds of significant column density.   
 This would result in a reduced \Hb\ emissivity but the Fe \Ka\ line will 
hardly be affected. 
 As already explained, the reddening correction for a dusty H{\sc ii} region 
can differ substantially from the correction procedure applied here. 

We have examined the much larger sample in Turner \etal\ (1997a) to estimate
the intrinsic I(Fe \Ka)/I(\Hb) ratio in type II AGN. 
 Out of 17 sources 
with reliable Fe \Ka\ and \Hb\ measurements (including the ones in this paper), 
 7 show  reddening corrected  
I(Fe \Ka)/I(\Hb)$<0.2$ which we consider consistent with origin in the NLR of 
these galaxies. 
 The remaining sources show a larger ratio that requires \Rf\ in excess of 
what is expected from the NLR gas.  
 The neutral Fe \Ka\ line in those is likely due to absorption by a larger 
column density, very neutral material that is either the walls of the 
central torus or large molecular clouds in the nucleus.

Given the likely origin of the 6.4 keV line, we can now comment on the nature
and location of the gas producing the high ionization iron lines. 
 Assuming the same ${\rm n_{Fe}/n_H}$ 
  in both components, we can derive  \Rf(hot)/\Rf(cold). 
 This is found to be about 1 for NGC~1068 and about 0.1 for Mkn~3 and Circinus. 
The uncertainty  is about a factor 2 since, as explained, the
scattering efficiency differs by about this factor when comparing Compton-thin
and Compton-thick gas.
 We  further consider possible combinations of covering factor and note that for
Compton-thin gas, 
 \Rf$\propto$\Ncol$\times$\Cf. This, combined with 
 \Ux(cold)/\Ux(hot)
(about 10$^{-3}$ with a large uncertainty,  see Fig. 1 and the parameters
used earlier for the NLR gas), enables us to estimate several likely
combinations of these quantities.

An interesting possibility involving the NLR idea, is that
 \Ncol(hot)$\sim 10^{-2}$\Ncol(cold).
 This would imply \Cf(hot)$\simeq 10$\Cf(cold), i.e.
\Cf(hot)$\simeq$1. In this case, the hot and cold components coexist, spatially,
and the large \Cf(hot) does not allow a torus with a small opening angle.
The typical NLR density is about $10^4$~\cc, 
thus  \nh(hot)$\sim 10 {\rm cm^{-3} }$.
The physical thickness of the hot gas in this case is of order 10-100 pc, i.e. of the
same order of the NLR size. We note, however, that the two components are
not in pressure equilibrium since
nkT$_{\rm e}$(cold)$\simeq 10$nkT$_{\rm e}$(hot). 
Another possibility is 
that \Cf$\sim 0.1$ in both components and \Ncol(cold)$\simeq 10$\Ncol(hot).
This does not allow a co-spatial existance of the two components. 
Finally, the 6.4 keV line may be from the thick torus walls. The efficiency
in this case is very large and suggests that the fraction of this wall visible
to us is extremely small. It also suggests that the hot gas completely fills the
opening in the torus.

Acknowledgments: It is a pleasure to acknowledge stimulating and useful discussions
with our colleagues R. Mushotzky, K. Nandra, T. Yaqoob and T. Kallman. 
A very useful referee report helped us improve the presentation of this paper.
This research is supported by the Universities Space Research Association 
(TJT, IMG) and by a special grant from the Israel Science Foundation (HN).

\begin{deluxetable}{lclcc}
\tablecaption{X-ray Emission Lines}
\tablehead{
\colhead{Line ID (Energy)}& \colhead{Line Flux} & \colhead{continuum}
& \colhead{EW} \\
\colhead{keV} & \colhead{$10^{-4}$ photons/cm$^2$/s}
& \colhead{$10^{-4}$ photons/cm$^2$/s/keV} & \colhead{Observed (eV)}
}
\startdata
\hline 
\multicolumn{4}{l}{Circinus} \nl
\hline 
\MgXI  & $0.21\pm0.08$ & 4.9 & 43 \nl
\MgXII  &$<0.10$ & 4.4 & $<23$ \nl
Si{\sc xii--xiii}  & $0.33^{+0.22}_{-0.08}$ & 3.4 & 100 \nl
\SiXIV\  & $<0.10$ & 3.0 & $<33$ \nl
S{\sc xiv--xv}  & $0.32\pm0.11$ & 2.4 & 133 \nl
\SXVI  & $<0.06$ & 2.1 & $<35$ \nl
Fe{\sc i--xxiv} & $3.10\pm0.19$ & 1.5 & 2067 \nl
Fe{\sc xxv--xxvi} + K$\beta$ & $0.86\pm0.05$ & 1.4& 614 \nl
\multicolumn{4}{l}{F(\Hb)=3$\times 10^{-12}$ erg/sec/cm$^2$} \nl
\multicolumn{4}{l}{ I(\Ha)/I(\Hb)=19}  \nl
\hline
\multicolumn{4}{l}{NGC 6240} \nl
\hline
\MgXI  & $<0.05$ & 1.80 & $<28$ \nl
\MgXII  &$<0.13$ & 1.60 & $<81$ \nl
Si{\sc xii--xiii} & $0.11^{+0.09}_{-0.04}$ &1.0 & 110 \nl
\SiXIV  & $<0.02$ & 0.70 & $<28$ \nl
S{\sc xiv--xv} & $<0.14$ & 0.60 & $<233$ \nl
\SXVI & $<0.05$ & 0.50 & $<100$ \nl
Fe{\sc i--xxiv} & $2.40^{+1.84}_{-0.76}$ & 2.65 & 906 \nl
Fe{\sc xxv--xxvi} + K$\beta$ & \nodata & \nodata & \nodata \nl
\multicolumn{4}{l}{F(\Hb) $>$ 5$\times 10^{-11}$ erg/sec/cm$^2$} \nl
\multicolumn{4}{l}{ I(\Ha)/I(\Hb)=63(?)}  \nl
\hline
\multicolumn{4}{l}{Mrk 3} \nl
\hline
\MgXI  & $<0.07$ & 1.60 & $<44$ \nl
\MgXII & $<0.12$ & 1.20 & $<100$ \nl
Si{\sc xii--xiii} & $0.11\pm0.04$ & 0.64 & $172$ \nl
\SiXIV  & $<0.01$ & 0.51 & $<20$ \nl
S{\sc xiv--xv} & $0.07^{+0.07}_{-0.03}$ & 0.39 & 179 \nl
\SXVI & $<0.04$ & 0.36 & $<111$ \nl
Fe{\sc i--xxiv} & $0.38\pm0.08$ & 0.40 & 950 \nl
Fe{\sc xxv--xxvi} + K$\beta$ & $0.07\pm0.01$ & 0.38 & 184 \nl
\multicolumn{4}{l}{F(\Hb)=4$\times 10^{-12}$ erg/sec/cm$^2$} \nl
\multicolumn{4}{l}{ I(\Ha)/I(\Hb)=6.6}  \nl
\hline
\tablebreak
\multicolumn{4}{l}{Mrk 348} \nl
\hline
\MgXI  & $<0.02$ & 0.20 & $<112$ \nl
\MgXII & $<0.01$ & 0.15 & $<44$ \nl
Si{\sc xii--xiii} & $<0.49$ & 0.10 & $<490$ \nl
\SiXIV  & $<0.04$ & 0.01 & $<214$ \nl
S{\sc xiv--xv} & $0.06\pm0.03$ & 0.10 & 600 \nl
\SXVI & $<0.01$ & 0.02 & $<68$ \nl
Fe{\sc i--xxiv} & $0.14\pm0.07$ & 1.24 & 113 \nl
Fe{\sc xxv--xxvi} + K$\beta$ & $0.05\pm0.05$ & 1.0 & 48 \nl
\multicolumn{4}{l}{F(\Hb)=8.4$\times 10^{-13}$ erg/sec/cm$^2$} \nl
\multicolumn{4}{l}{ I(\Ha)/I(\Hb)=7.4}  \nl
\hline
\multicolumn{4}{l}{NGC 4388} \nl
\hline
\MgXI  & $<0.04$ & 1.0 & $<40$ \nl
\MgXII & $<0.03$ & 0.80 & $<38$ \nl
Si{\sc xii--xiii} & $0.08^{+0.09}_{-0.04}$ & 0.50 & $154$ \nl
\SiXIV  & $0.05\pm0.04$ & 0.35 & $143$ \nl
S{\sc xiv--xv} & $0.10^{+0.08}_{-0.05}$ & 0.28 & 321 \nl
\SXVI & $<0.04$ & 0.26 & $<154$ \nl
Fe{\sc i--xxiv} & $0.70\pm0.39$ & 1.10 & 636 \nl
Fe{\sc xxv--xxvi} + K$\beta$ & $0.06\pm0.06$ & 0.78 & 77 \nl
\multicolumn{4}{l}{F(\Hb)=3.6$\times 10^{-13}$ erg/sec/cm$^2$} \nl
\multicolumn{4}{l}{ I(\Ha)/I(\Hb)=5.9}  \nl
\tablecomments{
 Line Energy (col. 1) and identification (col. 2). (f) denotes the
line energy was within 50 eV of a known line, and thus fixed at that
energy. Column 4: The Line width, $\sigma$, where this is not given, the
line was fixed at a width of 3 eV. Column 5: The observed line equivalent width.The tabulated \Hb\ fluxes are reddening corrected, as explained in the text.}
\enddata
\end{deluxetable}

\begin{table}[t]
\begin{minipage}{11cm}
\caption{Emperical continuum fit parameters}
\begin{tabular}{lccccc}
\hline
Object  & $\Gamma_{soft}$ & $\Gamma_{hard}$ & $N_{H,gal}$ (cm$^{-2}$) &
L$_{2-10}$\footnote{Observed Luminosity in ${\rm erg\, s}^{-1}$,
assuming H$_0=50$, q$_0=0.5$} &
 L(soft)=L(hard)\footnote{The energy at which the soft and the hard
components contribute  equal amounts} \\
\hline
Circinus &$1.84\pm 0.55$ &$-0.17\pm 0.35$&  $3.0 \times 10^{21}$ & $1.3 \times 10^{41}$ & 3.5 keV \\
NGC 6240\footnote{additional column of $1.5^{+1.4}_{-0.8} \times 10^{24}$
 cm$^{-2}$ covering $0.99^{+0.01}_{-0.07}$ of
the continuum is included in the model} &$1.94\pm0.13$
& $1.94\pm0.13$ & $5.5 \times 10^{20}$  & $5.2 \times 10^{42}$ & 5.7 keV\\
Mrk 3 & $2.90\pm0.51$ & $0.32\pm0.35$ & $8.7 \times 10^{20}$ &  $2.6 \times 10^{42}$ & 2.6 keV  \\
Mkn 348\footnote{additional column of $1.41 \times 10^{23}$ cm$^{-2}$
covering the hard component is included in the model} &
$2.77\pm0.87$ & $1.50\pm0.30$ & $1.2 \times 10^{20}$ &
$4.0 \times 10^{42}$ & 2.2 keV \\
NGC 4388  & $3.5\pm0.5$ & $-0.62^{+0.65}_{-0.21}$ & $2.6 \times 10^{20}$
& $ 2.0 \times 10^{42} $ & 2.8 keV \\
\hline
\end{tabular}
\end{minipage}
\end{table}

\begin{figure}
\plotfiddle{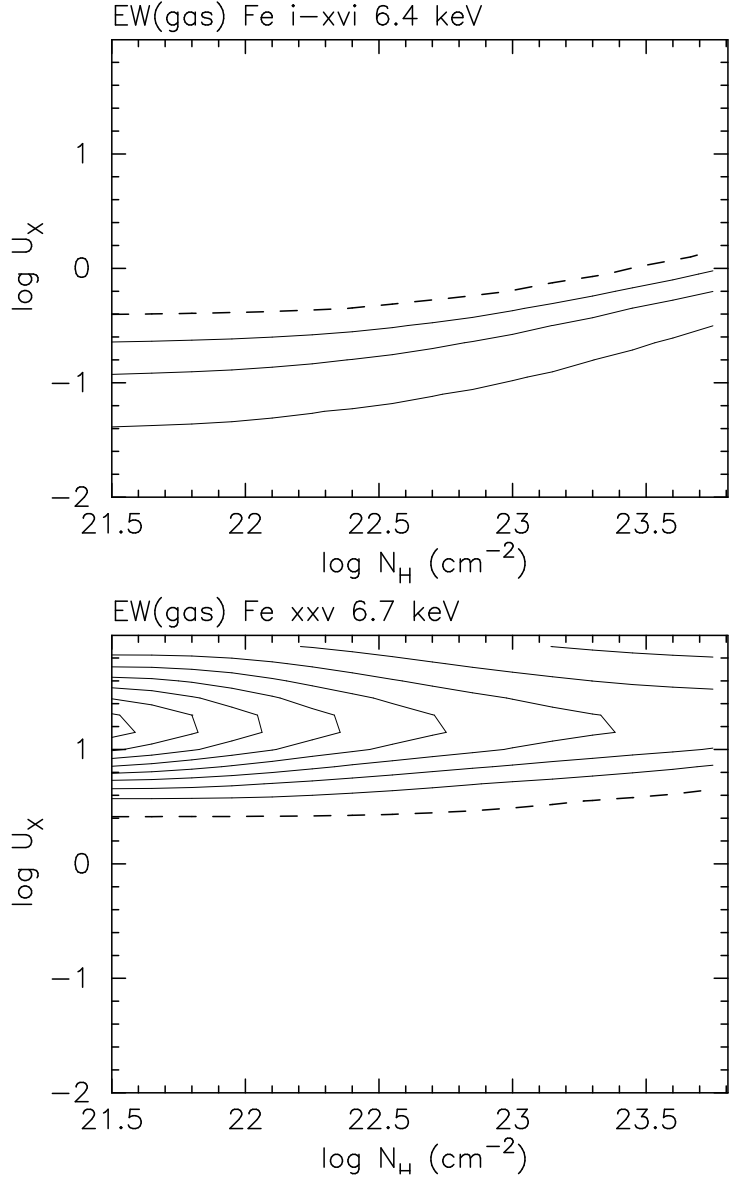}{13cm}{270}{80}{80}{-280}{460}
\figcaption{
 Calculated EW(Fe \Ka), relative to the scattered continuum,
 for   ``neutral''
 (Fe{\sc i--xvi})  and He-like iron lines. The 1--50 keV energy slope is 0.9, the
covering factor is 0.5, ${\rm n_{Fe}/n_H}=4 \times 10^{-5}$ and the microturbulent
velocity equals the local sound speed. The contours are separated by
200 eV, starting at EW(Fe \Ka)=100 eV (dashed line). Note the large EW of the
\FeXXV\ line at small columns due to resonance absorption.
The results are insensitive to the
exact covering fraction except for \Cf$>0.95$.
}
\end{figure}

\begin{figure}
\plotfiddle{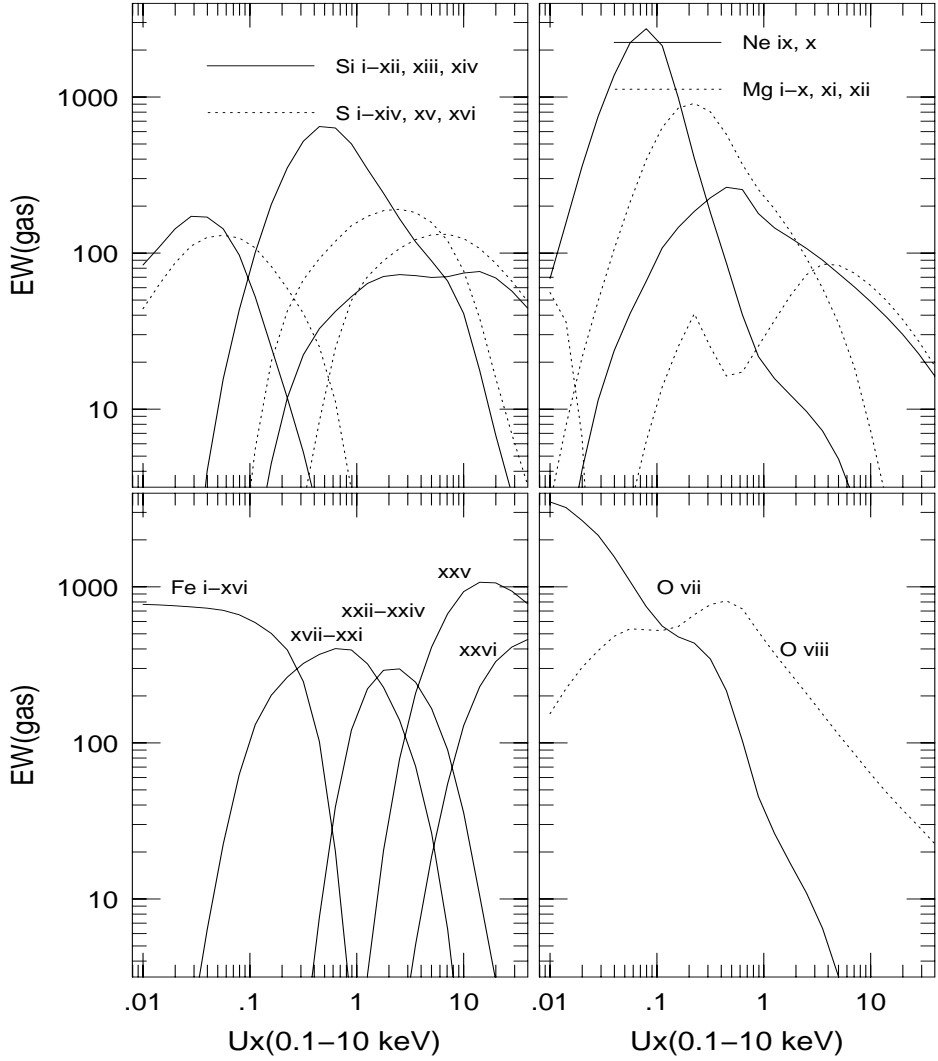}{15cm}{0}{70}{60}{-210}{20}
\figcaption{
 Equivalent widths, in eV,  relative to the scattered continuum, as a function of
ionization parameter.
 Standard composition gas (see text) with density of 10$^8$\cc, column
density of 10$^{22.4}$ \cmii, N$_{\rm E} \propto {\rm E}^{-1.9}$,
and  covering fraction of 0.5.
The top two panels show three lines for each of neon, magnesium,
silicon and sulphur; a ``neutral'' line (i.e. all transitions in ions
with three or more electrons), the He-like line and the H-like line.
The lines are easily recognized by the increasing level of ionization
(from left to right in the diagram).
}
\end{figure}
 
\begin{figure}
\plotfiddle{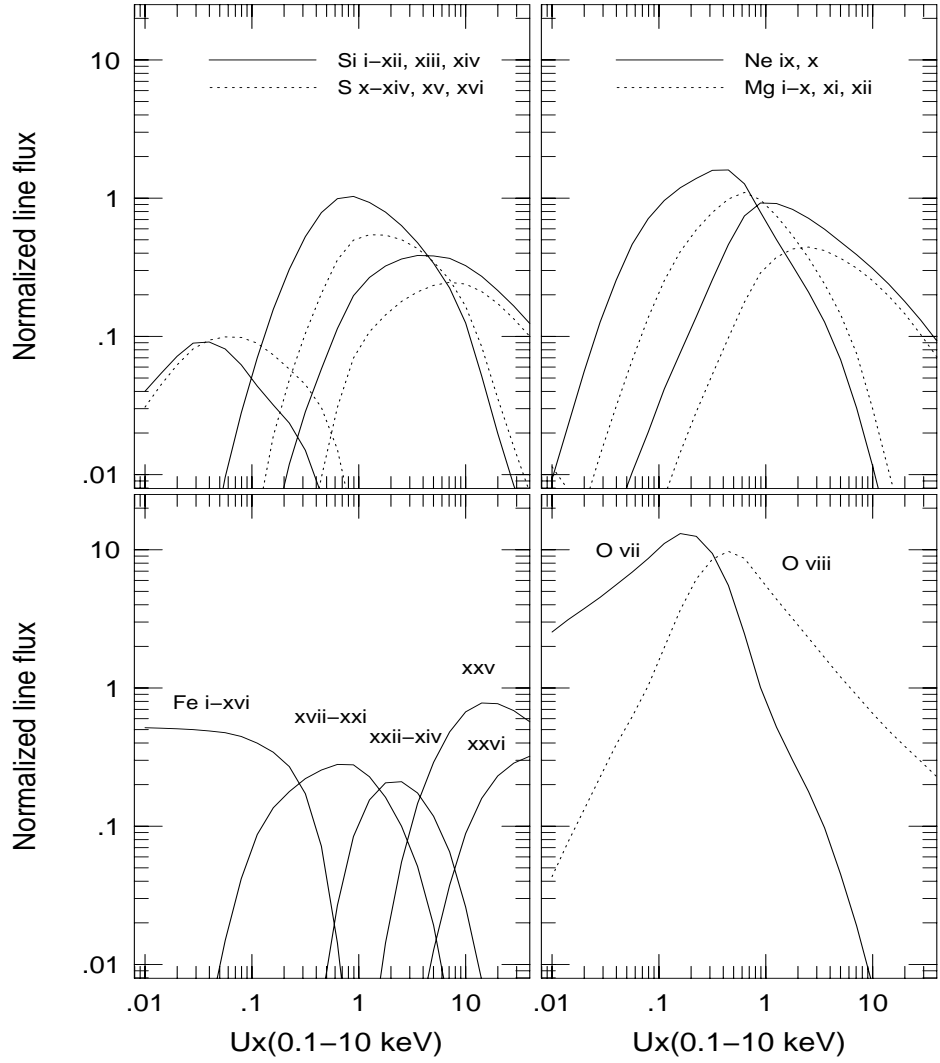}{15cm}{0}{70}{60}{-210}{20}
\figcaption{
As in Fig. 2 except that normalized line intensities, relative to
the incident continuum  at 0.65 keV, are shown.
}
\end{figure}

\begin{figure}
\plotfiddle{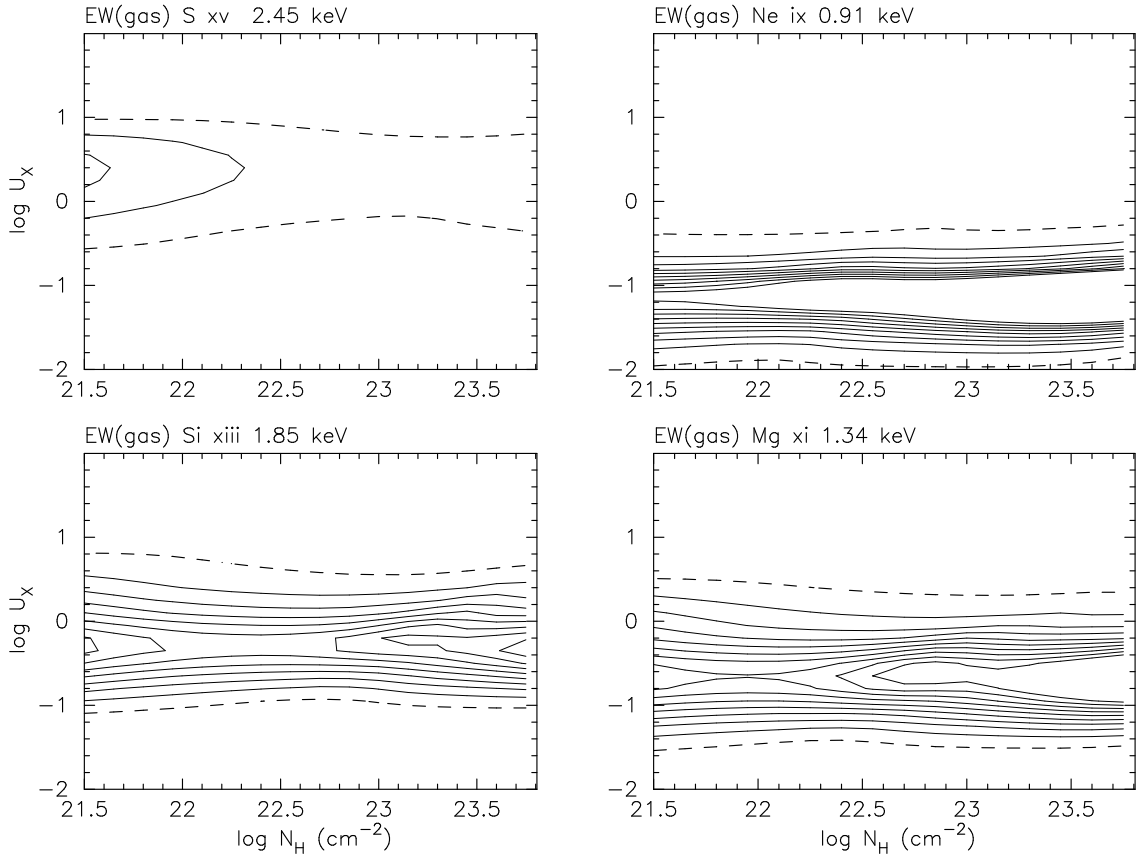}{10cm}{270}{60}{60}{-260}{460}
\figcaption{
Line equivalent widths, relative to the scattered
continuum, for various soft X-ray lines. Contours start at EW=100 eV and are separated
by 100 eV, except for \NeIX\ where they start at 100 eV but
 separated by 200 eV. The
lowest contour in all cases is plotted with a dashed line.
}
\end{figure}

\begin{figure}[t]
\plotfiddle{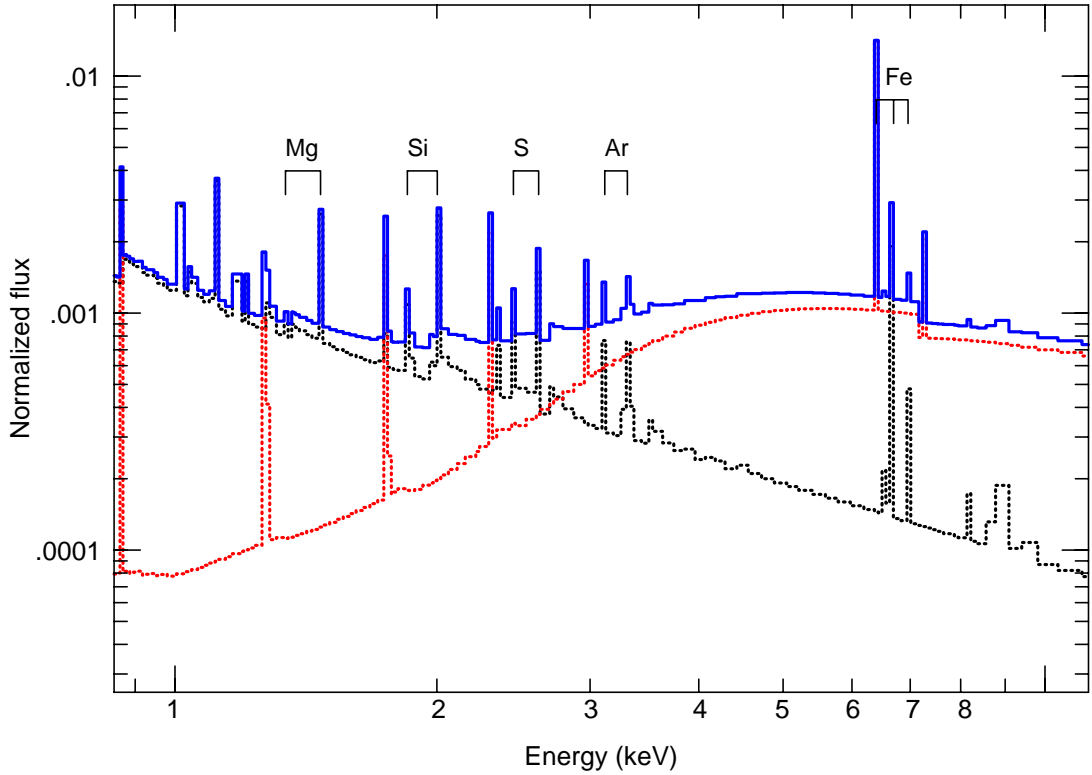}{6cm}{0}{60}{60}{-240}{30}
\figcaption{
 Scattered SEDs for an intrinsic
   N$_{\rm E} \propto {\rm E}^{-1.9}$ continuum and two different clouds:
one with \Ncol=10$^{23}$\cmii\ and Ux=$10^{-3}$ (the rising spectrum) and
the second with  \Ncol=10$^{22}$\cmii\ and Ux=10. The two are normalized to givea ratio of about 10:1 at 6.4 keV, which produces the Fe \Ka\ line ratios
observed in all cases except for NGC~1068. Note the very flat appearance of
the combined (solid line) spectrum).
}
\end{figure}

\begin{figure}[t]
\plotfiddle{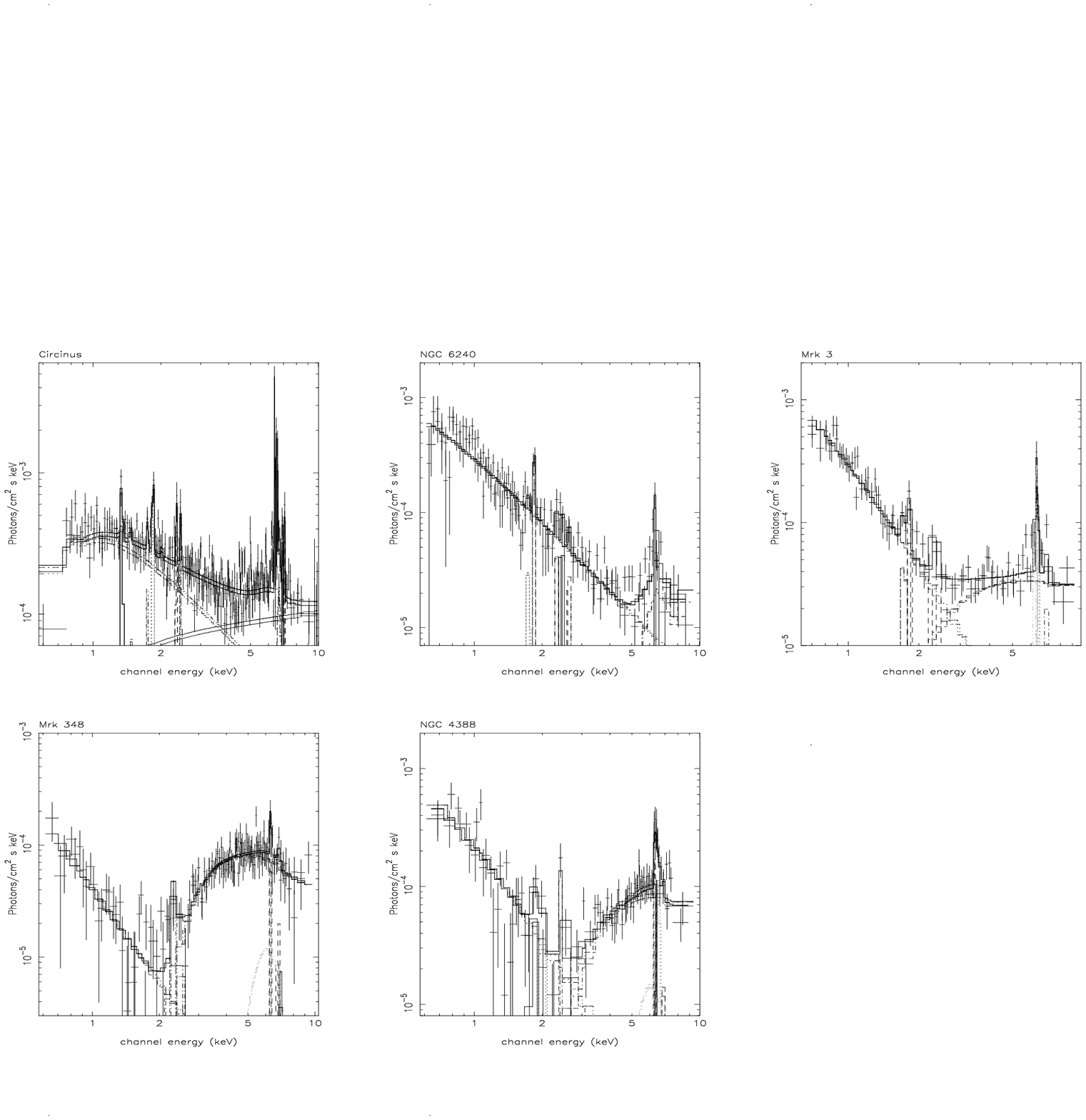}{15cm}{0}{30}{30}{-210}{-100}
\figcaption{
SIS data and fitted models for five scattering--dominated AGN.
}
\end{figure}

\begin{figure}
\plotfiddle{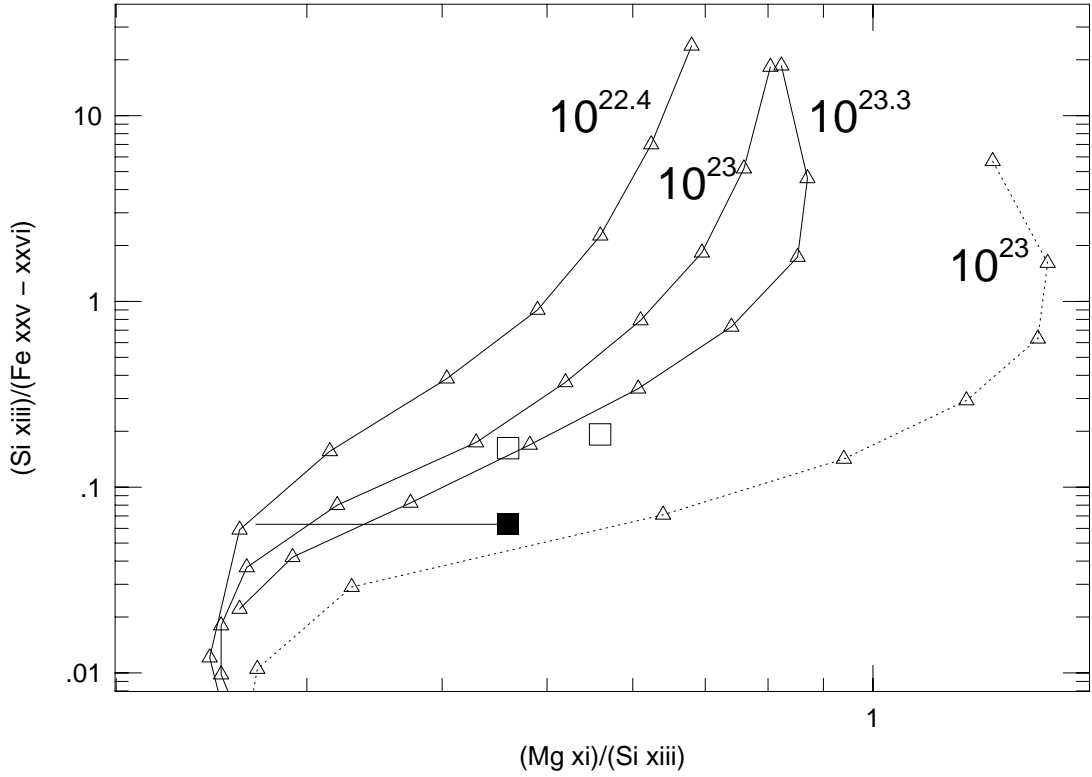}{8cm}{0}{60}{60}{-240}{30}
\figcaption{
A line ratio diagram for three AGN. The curves
   represent series of  increasing ionization parameter for the column density
   marked. In all curves, Ux=10$^{-0.4}$ at the top and increases downward,
 by 0.1dex per step. The solid lines are for the assumed standard
 composition and the dotted line for
   n$_{\rm Fe}$/\nh\ three times larger and all other abundances not altered.
}
\end{figure}

\begin{figure}
\plotfiddle{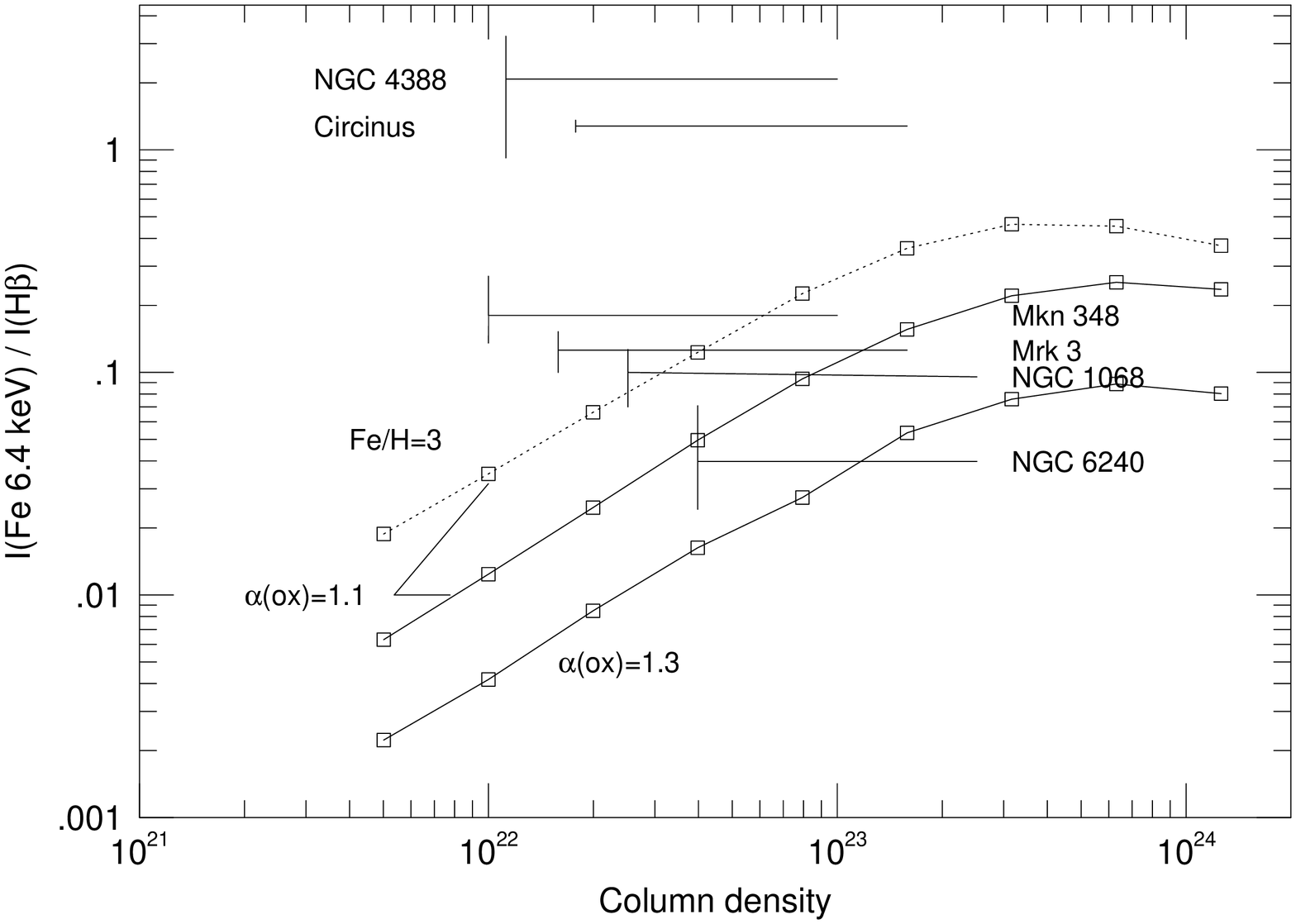}{8cm}{0}{50}{50}{-220}{50}
\figcaption{
Theoretical Fe \Ka/\Hb\ as a function of column density
   for the L$_{\rm E} \propto {\rm E}^{-0.9}$ X-ray
   continuum,  \Cf=0.1, Ux=$10^{-3}$ and various \aox, as marked.
   The solid lines are for the standard metallicity
    and the dotted line for n$_{\rm Fe}$/\nh\ three times larger.
  The horizontal lines mark the observed line ratio in our sample. The
associated error bars correspond to the uncertainties in the Fe \Ka\
fluxes and do not includ  the uncertain reddening correction factors.
}
\end{figure}

\end{document}